\documentclass[twocolumn,showpacs,pra,amsmath,amssymb,floatfix]{revtex4-1}
\input{epsf}

\usepackage{graphicx}
\usepackage{dcolumn}
\usepackage{bm}
\usepackage{amssymb}

\begin{document}
\title{Non-equilibrium dynamics of spin-orbit coupled 
lattice bosons}
\author{H. T. Ng}
\affiliation{Center for Quantum Information, Institute for Interdisciplinary Information Sciences, Tsinghua University, Beijing 100084, P. R. China}

\date{\today}

\begin{abstract}
We study the non-equilibrium dynamics of two component
bosonic atoms in a one-dimensional optical lattice in
the presence of spin-orbit coupling. In the Mott insulating regime, the two-component bosonic
system at unity filling can be described by
the quantum spin XXZ model. The atoms are initially prepared in their lower spin states. The system becomes out of equilibrium by suddenly introducing spin-orbit coupling to the atoms.  The system shows the relaxation and non-stationary dynamics, respectively, in the different interaction regimes. We find that the
time average of magnetization is useful to characterize
the many-body dynamics. The effects of even and 
odd numbers of sites are discussed. Our result sheds light on non-equilibrium dynamics due to the interplay between spin-orbit coupling and atomic interactions.
\end{abstract}


\maketitle

\section{Introduction}
Non-equilibrium dynamics of a closed many-body system
is of fundamental importance \cite{Polkovnikov,Eisert} in quantum physics and statistical mechanics. For example, the local observables appear to be in thermal states even if the entire many-body system is in a pure state. The eigenstate thermalization hypothesis \cite{Deutsch,Srednicki,Rigol} was proposed to explain such phenomena for complex quantum systems. However, it cannot be applied to integrable
systems. In fact, a general mechanism of thermalization
is still lacking \cite{Eisert}. It is important to study the dynamics of an isolated many-body system in experiments. This may help to understand the behaviour of non-equilibrium dynamics and thermalization mechanism.

Ultracold atoms offer experimental platforms to
study many-body dynamics of closed systems \cite{Langen}. 
For instance, ultracold atoms can provide
a relatively long accessible time to study the non-equilibrium dynamics, and the high level of controllability to adjust the interaction parameters
for initiating the many-body dynamics.
In addition, ultracold atoms have been
exploited to simulate a lot of intriguing quantum
phenomena such as quantum phase transition \cite{Greiner} from a Mott insulating regime to superfluid. The advanced detection techniques have been invented to enable one to
individually address a single atom \cite{Bakr,Weitenberg}. Therefore,
this can be used for probing the dynamics in the microscopic description \cite{Fukuhara}. Recently, the relaxation dynamics of closed ultracold atomic systems have been observed \cite{Trotzky,Gring}.

In this paper, we consider a system of two-component
bosonic atoms in a one-dimensional (1D) optical lattice.
In the Mott insulating regime, the system, with unit filling, can be described by a quantum spin XXZ model \cite{Schollwock}. We consider all atoms to be initially prepared in their lower spin states. To study the non-equilibrium dynamics,
the spin-orbit (SO) coupling \cite{Galitski} is suddenly
turned on. In fact, SO coupling, which produces the interaction between the particle's spin and momentum, naturally exists in solid-state
materials. It gives rise to a number of intriguing
effects such as topological
insulators and superconductors, etc. \cite{Hasan,Qi}.
Spin-orbit coupling in atomic Bose-Einstein condensates \cite{Lin} can be produced by inducing two-photon Raman transition using a pair of lasers \cite{Liu0}.   
Alternatively, the SO coupling between the atoms in the lattice can be induced by periodically shaking
the lattice potential \cite{Sengstock,Goldman}.
More recently, the techniques for adjusting SO coupling have been shown \cite{Garcia,Luo}. The SO coupling gives rise to pair interactions between two neighbouring atoms and Dzyaloshinskii-Moriya (DM) interactions \cite{Radic,Cole,Piraud,Zhao,Ng} in the lattice.
The DM interaction leads to rich magnetic phase diagram \cite{Cole}, for example, it can induce spin spirals.

We consider the magnetization as an observable to
study the many-body dynamics. We find that the time average of magnetization is useful for characterizing the non-equilibrium dynamics.
The system exhibits the relaxation and non-stationary dynamics in the different interaction regimes
which depend on the SO coupling strength and
the ratio of inter-component interaction to intra-component interaction. Indeed, the dynamical behaviours
relate to the overlap between the initial state
and eigenstates. The interplay between the SO coupling and atomic interactions leads to the changes of this overlap, and results in the different dynamical
behaviours.

Thermalization \cite{Goldstein,Popescu,Genway,Banuls}
occurs when a subsystem evolves to a mixed state even if the system is in a pure state. 
In the relaxation regime, the spins rapidly relax just after the SO coupling is turned on. The degree of quantum coherence of local spins can be measured by using the purity. An atom in each site will evolve to a nearly completely mixed state in a sufficiently long time.

The transition from the relaxation dynamics to
non-stationary evolution occurs when the SO coupling strength is strong and the inter-component
interaction strength becomes sufficiently larger than the intra-component interaction. In the non-stationary regime, we find that the distinct dynamics are displayed for the even and odd numbers of sites, respectively. For even-number cases, the effective two-level dynamics is shown. This forms a superposition of two distinct states \cite{Leggett} during the time evolution. These superposition states are useful for quantum metrology \cite{Giovannetti}. On the other hand, the spin system becomes ferromagnetic in the odd-number cases.

This paper is organized as follows: In Sec.~II, we introduce the system. In Sec.~III, we study the non-equilibrium dynamics of this system. We characterize the many-body
dynamics by using the time average of magnetization.
We discuss the relaxation dynamics of local spins
and non-stationary dynamics in the different interaction regimes. The effects of even and odd number of sites on the dynamics are discussed. We provide a discussion and a conclusion in the sections IV and V, respectively. In Appendix A, we derive the effective Rabi frequency.

\section{System}
We consider the two-component bosonic atoms to be
trapped in a one-dimensional optical lattice. We assume
that this system has open boundary
conditions.
The two-component Bose-Hubbard model can
be used to describe the interactions of 
two-component bosons in an optical lattice.
The Hamiltonian $H_{\rm BH}$ can be written as, ($\hbar=1$),
\begin{eqnarray}
H_{\rm BH}&=&\sum_{\alpha,i}\Big[J_\alpha(\alpha^\dag_i\alpha_{i+1}+{\rm H.c.})
+\frac{U_\alpha}{2}{n}^\alpha_i(n^\alpha_i-1)\nonumber\\
&&+U_{ab}n^a_in^b_i\Big],
\end{eqnarray}
where $\alpha_i$ and $\alpha^\dag_i$ are the annihilation
and creation operators of an atom in the atomic spin state $|\alpha\rangle$, 
and $n^\alpha_i$ is the number operator at site $i$, and $\alpha=a,b$.
The parameter $J_\alpha$ is the tunnel coupling,
$U_{a(b)}$ and $U_{ab}$ are the intra- and inter-component interaction strengths of atoms, respectively.
We assume that the tunnel coupling and atom-atom interaction strength of each component are nearly equal, i.e., $J_a{\approx}J_b{\approx}J$ and
$U_a{\approx}U_{b}{\approx}{U}$. 

We consider that the atom-atom
interaction strengths ${U}$ and $U_{ab}$ are repulsive,
and also they are much larger than the parameters such as $J$ and $t_{so}$. In this strongly interacting regime with unit filling, it is convenient to write the two-mode bosonic operators in terms of angular momentum operators, i.e., $S^+_{i}=a^\dag_ib_i$, $S^-_{i}=b^\dag_i{a_i}$ and $S^z_{i}=(b^\dag_i{b_i}-a^\dag_i{a}_i)/2$. The Hamiltonian can be written in terms of spin operators as:
\begin{eqnarray}
\label{XXZ}
H_{\rm XXZ}\!&=&\!\lambda\!\sum^{N-1}_{i=1}\!\Big[2(1-2U_r)S^z_iS^z_{i+1}\!-\!(S^+_iS^-_{i+1}+S^-_{i}S^{+}_{i+1})\Big],\nonumber\\
\end{eqnarray}
where $\lambda=2J^2/U_{ab}$ and $U_r=U_{ab}/U$. The system can be described by the quantum XXZ model \cite{Duan}.

To study the non-equilibrium dynamics, we consider the SO coupling to be suddenly applied to the atoms. For a non-interacting single-particle Hamiltonian, it is given by
\begin{eqnarray}
H^s_{\rm SO}&=&\frac{\mathbf{k}^2}{2m}I+\beta{k_x}\sigma_y+\delta{k_y}\sigma_x,
\end{eqnarray}
where $\mathbf{k}=(k_x,k_y,k_z)$ is the momentum of a particle with a mass $m$, $\beta$ and $\delta$ are the SO coupling strengths, and $I$ and $\sigma_{x,y}$ are the identity and the Pauli spin operators, respectively. If $\delta$ equals $-\beta$, then it is called
as the Rashba SO coupling \cite{Lin}. The Dresselhaus SO coupling \cite{Lin} is an alternative form of spin-orbit coupling where $\beta$ equals $\delta$ and they are both negative \cite{Lin}. Recently, the SO coupling with an equal weight of Rashba and Dresselhaus couplings has been realized in a ${}^{87}$Rb Bose-Einstein condensate \cite{Lin}, i.e.,
$\beta\neq{0}$ and $\delta=0$. By inducing a two-photon
Raman transition via the laser beams, the equal weight of Rashba and Dresselhaus SO couplings can be produced, $\propto{k_x\sigma_y}$.
The Hamiltonian, which describes such SO coupling in a 1D optical lattice, can be written as \cite{Liu,Cai}
\begin{eqnarray}
H_{\rm SO}&=&t_{so}\sum_{i}(a^\dag_ib_{i+1}-a^\dag_ib_{i-1}+{\rm H.c.}),
\end{eqnarray}
where $t_{so}$ is the strength of spin-orbit coupling.

In the presence of SO coupling,
the effective Hamiltonian is written as \cite{Piraud,Zhao}
\begin{eqnarray}
\label{effsHam}
H^s_{\rm eff}&=&\lambda\sum^{N-1}_{i=1}\Big\{2\Big[\Big(\frac{t_{so}}{J}\Big)^2-1\Big]\Big(2U_r-1\Big)S^z_iS^z_{i+1}\nonumber\\
&&\!\!+\Big(\frac{t_{so}}{J}\Big)^2(S^+_iS^+_{i+1}+S^{-}_iS^-_{i+1})\!-\!(S^+_iS^-_{i+1}+S^-_{i}S^{+}_{i+1})\nonumber\\
&&-4\frac{U_{ab}}{U}\Big(\frac{t_{so}}{J}\Big)(S^z_iS^x_{i+1}-S^x_iS^z_{i+1})\Big\}.
\end{eqnarray}
The last terms in Eq.~(\ref{effsHam}) are called
the DM interactions \cite{Piraud,Zhao}. Now the total system can be described by the XYZ spin model with the DM interactions \cite{Piraud,Zhao,Ng}. 

Let us briefly discuss the various terms in the effective Hamiltonian in Eq.~(\ref{effsHam}). The first term $S^z_iS^z_{i+1}$ favors to preserve the same polarization of two neighboring spins with their initial states. This gives rise to bound magnons in the XXZ chain \cite{Fukuhara}. The second
term $S^+_iS^+_{i+1}+S^-_iS^{-}_{i+1}$ describes the interaction which excites and de-excites the two neighboring spins in pair simultaneously. The third term
$S^+_iS^-_{i+1}+S^-_{i}S^+_{i+1}$ leads to spin-exchange between two nearest neighbors. The forth term 
$S^z_iS^x_{i+1}-S^x_iS^z_{i+1}$ causes spin-rotation in which the rotation direction depends on the spin states
of their nearest neighbors. 

When $t_{so}/J$ and $U_r$ are roughly equal to one, all terms equally contribute to the dynamics. All the terms are in competition. This results in eigenstates with the different combinations of spin states.
On the contrary, when both parameters $t_{so}/J$ and $U_r$ are large, the first term becomes dominant. All spins tend to have the same polarization. This leads to the very different dynamics in these two interaction regimes.

\section{Many-body dynamics}
We investigate the quantum dynamics of the system by suddenly applying the SO coupling to the atoms. Initially, all atoms are prepared in their lower spin states, i.e., 
\begin{eqnarray}
\label{initial_state}
|\Psi(0)\rangle&=&|\downarrow\downarrow\ldots\downarrow\downarrow\rangle.
\end{eqnarray}
This state is indeed an eigenstate in Eq.~(\ref{XXZ}) for
$t_{so}=0$. To initiate non-equilibrium dynamics, a quantum quench has to be introduced. By suddenly turning on the SO coupling, the system is described by the Hamiltonian in Eq.~(\ref{effsHam}). The initial state is no longer an eigenstate of the Hamiltonian of the system.
If the SO coupling is suddenly turned on, then the system
becomes out of equilibrium.  
We numerically simulate the dynamics of this spin
chain by using exact diagonalization (see \cite{Schachenmayer} and references therein).

\begin{figure}[ht]
\centering
\includegraphics[height=10cm]{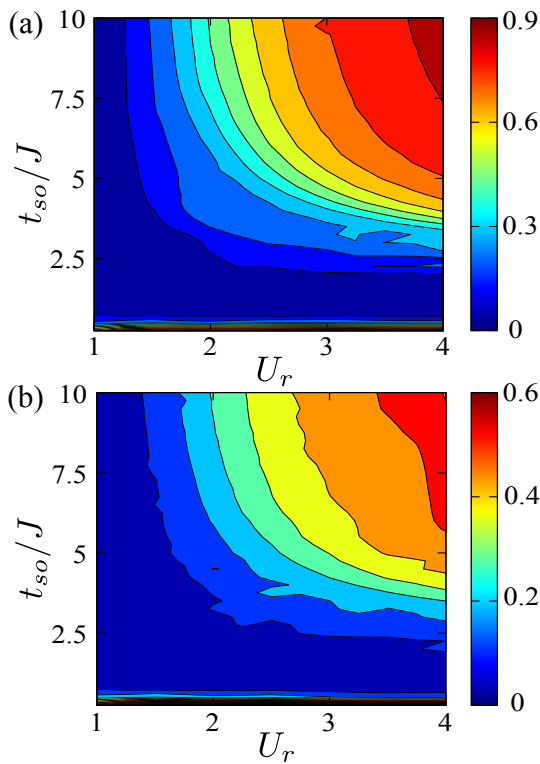}
\caption{ \label{contour_magnetization} (Color online) 
Contour plot of the average of magnetization $|M|$ versus parameters $t_{so}$ and $U_r$, where the total time taking for average is $\tau=10\lambda$.
The system's sizes $N=11$ and $N=12$ are shown in (a)
and (b), respectively.
}
\end{figure}

We study the dynamics of magnetization which is given by
\begin{eqnarray}
\label{magnetization}
M&=&\frac{2}{N}\sum^N_{i=1}\langle{S}^z_i\rangle.
\end{eqnarray}
The magnetization $M$ is equal to $+1(-1)$ when all
spins are in up(down) states. We take the average of
magnetization $|M|$ within a period $\tau=10\lambda$, i.e.,
\begin{eqnarray}
\label{bar_M}
\bar{M}=\frac{1}{\tau}\int^{\tau}_0{|M(t)|}dt.
\end{eqnarray}
By taking the absolute sign of the magnetization, 
we can ensure that $\bar{M}$ is positive. Here we set the period $\tau$ to be $10\lambda$. This period is sufficiently long until the local spins become steady. It enables us to characterize the dynamics by using 
$\bar{M}$ in Eq.~(\ref{bar_M}). 

In Fig.~\ref{contour_magnetization}, we plot the time average of magnetization $\bar{M}$ versus the parameters 
$t_{so}$ and $U_r$. When $t_{so}/J$ ranges from 0.75 to 2, $\bar{M}$ is close to zero. Note that the initial magnetization is -1. After taking the average, it becomes nearly zero.
This means that the system reaches a steady state and
the magnitude of magnetization is small in a long time.

As both $t_{so}$ and $U_r$ increase, $\bar{M}$ gradually increases. The system is no longer stationary in
a long time. This shows the different layers in Fig.~\ref{contour_magnetization}. When both parameters $t_{so}/J$ and $U_r$ become sufficiently large, $\bar{M}$ can reach nearly $0.5$
and 1, respectively, for the even and odd numbers of sites. In this regime, this shows that the different behaviours of even- and odd-number of sites.

It should be noted that 
the average magnetization is close to one in Fig.~\ref{contour_magnetization} if $t_{so}/J$
is less than 0.75. In fact, we cannot characterize 
the dynamics of this interaction range because the
time evolution is too slow and they do not reach the
steady state within the period $\tau=10\lambda$.

According to Fig.~\ref{contour_magnetization}, the many-body dynamics can be mainly classified into the two different types which are
relaxation dynamics and non-stationary evolution, respectively. We will discuss them in the following
subsections.

\subsection{Relaxation dynamics}
\begin{figure}[ht]
\centering
\includegraphics[height=8.5cm]{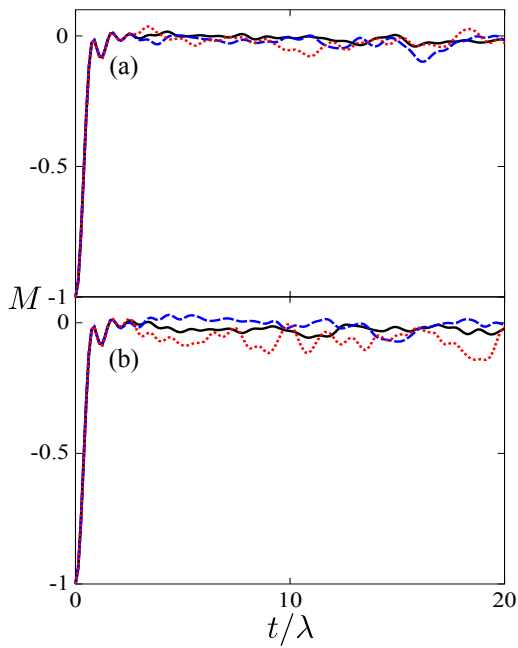}
\caption{ \label{relax_magnet} (Color online) 
Magnetization $M$ versus time $t/\lambda$, for $t_{so}=J$ and
$U_r=1$. The odd and even numbers of sites are shown
in (a) and (b). The different sizes of the system are denoted with the different lines: $N=11$ and 12 (red dotted), $N=13$ and 14 (blue dashed) and $N=15$ and 16 (black solid), respectively.
}
\end{figure}

Now we study the quantum dynamics of local spins in the parameter region which
the long-time average of magnetization $\bar{M}$ is about zero in Fig.~\ref{contour_magnetization}. In Figs.~\ref{relax_magnet}(a) and (b), we plot the dynamics of magnetization $M$, for the odd and even numbers of sites, respectively. Initially, the magnetization $M$ is equal to $-1$. When the SO coupling is turned on, the magnetization swiftly increases in a short time. Then, it becomes saturated to around zero in a longer time. This shows that the spins nearly relax to the steady states. As the system size increases, the magnetization becomes more steady. Besides, the relaxation dynamics shows no different in the odd- and even-number cases.

\begin{figure}[ht]
\centering
\includegraphics[height=8cm]{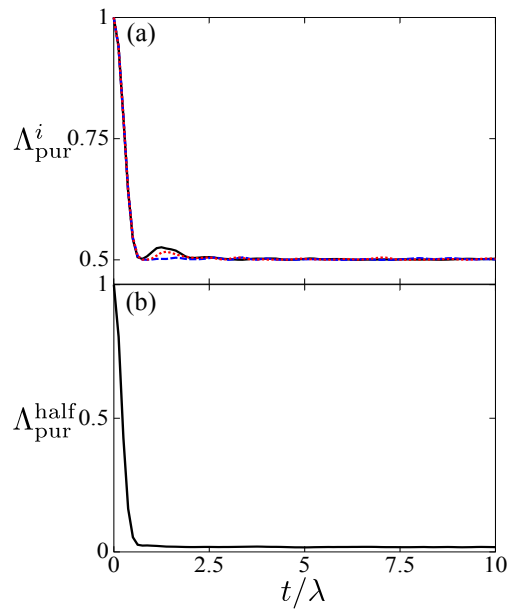}
\caption{ \label{purity} (Color online) 
Purities of local spins and half of a spin chain
versus time $t/\lambda$. In (a), the purity of
local spin $i$ versus dimensionless time $t/\lambda$
is shown. The different lines are
denoted for the purities of different ion $i$: $i=4$ (solid black), $i=6$ (blue dashed) and $i=8$ (red dotted), respectively. 
In (b), the purity of the left part of a spin chain versus time is plotted. The parameters are used: $N=12$, $t_{so}=J$ and $U_r=1$. 
}
\end{figure}

To study the relaxation of local spins, we examine their purities. The purity
is a quantity which measures the degree of quantum coherence of a system. It is defined as
\begin{eqnarray}
\Lambda_{\rm pur}={\rm Tr}(\rho^2),
\end{eqnarray}
where $\rho$ is the density matrix of a system.
If the system is in a pure state, then the purity
is equal to one. Otherwise, the purity is less than
one.

We investigate the purity $\Lambda^i_{\rm pur}$ of ion $i$, where $\Lambda^i_{\rm pur}={\rm Tr}(\rho^2_i)$ and $\rho_i$ is the reduced density matrix of ion $i$. The reduced density matrix $\rho_i$ can be obtained by tracing out the rest of the other spins in the chain.  
In Fig.~\ref{purity}(a), we plot
the purities of local spins versus time.
The initial purity is equal to 1. Afterwards, the purities rapidly drop and the purity $\Lambda^i_{\rm pur}$ of a local spin decreases to about 0.5 in a long time. 
The purity of a spin-half particle in a completely
mixed state is 0.5, where all diagonal elements of $\rho_i$ are equally weighted and off-diagonal elements are zero. This implies that the spins
are nearly fully relaxed.
Additionally, we study the purity $\Lambda^{\rm half}_{\rm pur}={\rm Tr}(\rho^2_{\rm half})$ of a half of the spin chain in Fig.~\ref{purity}(b), where the reduced density matrix
$\rho_{\rm half}$ can be obtained by tracing out another half of the chain. The purity becomes saturated to a value  0.018. It is very close to the totally mixed state 
which gives the purity $\tilde{\Lambda}^{\rm half}_{\rm pur}=2^{N/2-N}\approx{0.015625}$ in our case. This means
that the half of a spin chain can be approximately described by a totally mixed state.

\begin{figure}[ht]
\centering
\includegraphics[height=9cm]{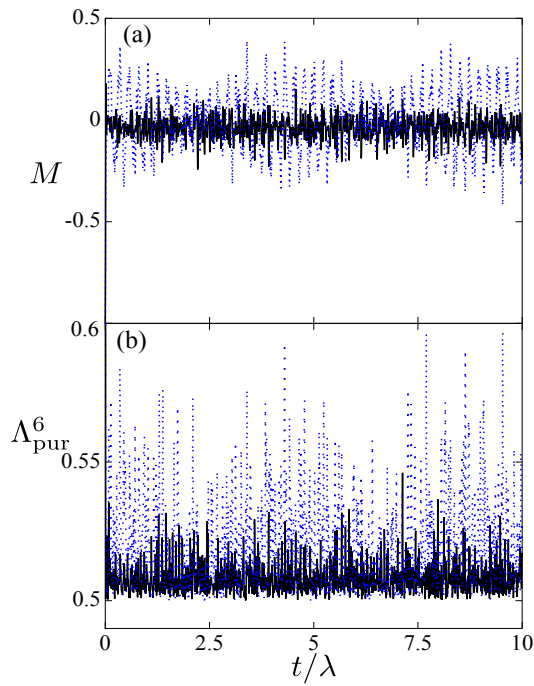}
\caption{ \label{relax_msz_purity} (Color online) 
Magnetization $M$ and purity $\Lambda^6_{\rm pur}$ of the 6-th spin versus time $t/\lambda$ are plotted in (a) and
(b), respectively, for $N=12$ and $t_{so}=8J$. The different parameters $U_r=1$ and $U_r=1.5$ are denoted with the black-solid and blue-dotted lines, respectively.
}
\end{figure}

Then, we compare the relaxation dynamics
with the different strengths of parameters $U_r$. The average magnetization $\bar{M}$ increases when $U_r$ becomes larger.
In Fig.~\ref{relax_msz_purity}(a),
we plot the magnetization $M$ as a function of time, for $U_r=1$ and 1.5, respectively. The magnetization $M$ in Eq.~(\ref{magnetization}) fluctuates around zero with a larger magnitude if $U_r$ becomes larger. Therefore, this will give a larger value of $\bar{M}$ in Eq.~(\ref{bar_M}) which is obtained by taking average of the absolute value of $M$.
We study the purity
of a local spin in Fig.~\ref{relax_msz_purity}(b). The purity increases when $U_r$ increases. This means that this local spin has a higher degree of quantum coherence. 
This suggests that $\bar{M}$ is an useful quantity to
characterize the relaxation dynamics of this system.

\begin{figure}[ht]
\centering
\includegraphics[height=11cm]{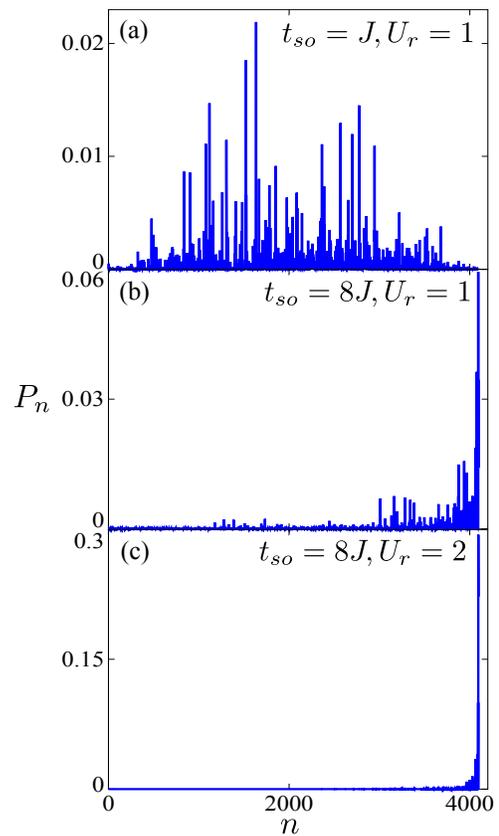}
\caption{ \label{overlap_prob} (Color online) 
Overlap probability $P_n$ versus the index $n$ of the $n$-th eigenstate, for $N=12$. In (a), the parameters
$t_{so}=J$ and $U_r=1$ are used. The different parameters $U_r=1$ and 2 are used in (b) and (c), respectively, but with the same SO coupling, $t_{so}=8J$.
}
\end{figure}
Indeed, the occurrence of relaxation can be 
understood by examining the overlap between the initial state and eigenvectors of the system. We consider the probability coefficients of the initial state and eigenvectors 
\begin{eqnarray}
P_n&=&|\langle\Psi(0)|E_n\rangle|^2,
\end{eqnarray}
where $|\Psi(0)\rangle$ and $|E_n\rangle$
are the initial state and the $n$-th eigenvectors.

In Fig.~\ref{overlap_prob}(a), we plot the overlap
probabilities $P_n$ versus $n$, where $t_{so}/J=U_r=1$  and $n$ is an index of the $n$-th eigenstate. This corresponds to the previous case
in Fig.~\ref{relax_magnet}. We can see that the initial
state has a large overlap with the eigenstates. The initial state overlaps
with almost entire eigen-spectrum.
In Fig.~\ref{overlap_prob}(b), the overlap probabilities
$P_n$ are plotted versus $n$. It corresponds to the case in Fig.~\ref{relax_msz_purity}, where the magnetization shows stronger fluctuations in the dynamics. Obviously, the overlap between the initial state and the eigenstates is much smaller than that in Fig~\ref{overlap_prob}(a). 
Here the off-diagonal terms of the observables are suppressed if there is a large overlap between the initial states and the eigenstates of the system.

\subsection{Even-odd effect}
When both parameters $t_{so}/J$ and $U_r$ are sufficiently large,
the dynamics of the system becomes non-stationary.
We find that the dynamical behaviours are totally
different between the even and odd numbers of sites. For even-number cases, the system undergoes an effective two-level dynamics. In contrast, the system becomes ferromagnetic if the number of sites is odd.

\subsubsection{Even number case: Effective two-level dynamics}
\begin{figure}[ht]
\centering
\includegraphics[height=9cm]{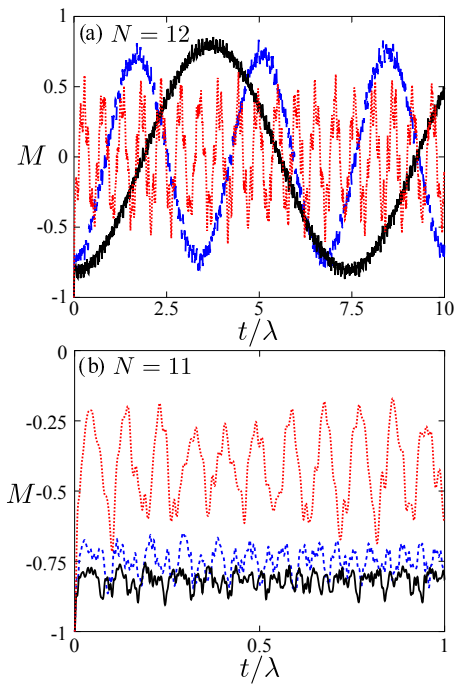}
\caption{ \label{even_odd} (Color online) 
Magnetization $\tilde{M}$
versus time $t/\lambda$ for even ($N=12$) and odd $(N=11)$ numbers in (a) and (b), respectively, and $t_{so}=8J$.
The different lines for the different $U_r$ are shown: $U_r=4$ (black solid line), $U_r=3$ (blue dash line)
and $U_r=2$ (red dotted line), respectively. 
}
\end{figure}

When $t_{so}/J$ and $U_{r}$ are both larger than one, the first term in the Hamiltonian in Eq.~(\ref{effsHam})
becomes dominant, i.e., $H_0\propto\sum_iS^z_{i}S^z_{i+1}$ and the other terms in the Hamiltonian are perturbations. Obviously, the states $(|\uparrow\uparrow\ldots\uparrow\uparrow\rangle\pm|\downarrow\downarrow\ldots\downarrow\downarrow\rangle)/\sqrt{2}$ are two nearly degenerate eigenstates of the Hamiltonian $H_0$. 
In this regime, the entire many-body dynamics can be effectively described by these two degenerate states if the system starts with
$|\downarrow\downarrow\ldots\downarrow\downarrow\rangle$
in Eq.~(\ref{initial_state}). 
We plot the time evolution of the magnetization in Fig.~\ref{even_odd}(a).
The magnetization shows periodic oscillations. The effective Rabi frequency decreases as the parameter $U_r$ increases and a larger magnitude can be attained. In fact, the superposition of the two degenerate ground states can be produced, i.e.,
\begin{eqnarray}
|\Psi(t)\rangle&\approx&c_1|\uparrow\uparrow\ldots\uparrow\uparrow\rangle+c_2|\downarrow\downarrow\ldots\downarrow\downarrow\rangle,
\end{eqnarray} 
where $|c_1|^2+|c_2|^2=1$.

The two degenerate states can be coupled via the high-order virtual transitions. The effective Rabi frequency can be approximately obtained which can be derived by using the high-order perturbation theory in Appendix A.
Since this effective Rabi frequency inversely scales with the power $N$, the rate of evolution becomes slow as the system's size increases. This hinders the creation of superposition of two spin states when the system goes large.

In Fig.~\ref{overlap_prob}(c), we plot the overlap between the initial state in Eq.~(\ref{initial_state}) and the eigenstates. The overlap is much smaller than
the two previous cases in Figs.~\ref{overlap_prob}(a)
and (b) which shows the relaxation dynamics. We have presumed that the system is strictly contained in the degenerate subspace which can be described by the two degenerate states only. Therefore, there are only two degenerate states involved in the entire dynamics. 
In the limit of strong interaction, the dynamics cannot be thermalized.

\subsubsection{Odd number case: Ferromagnetic}
In Fig.~\ref{even_odd}(b), we plot the magnetization
versus time for odd-number cases.
The spins tend to maintain in their ground states when
$U_r$ increases. The spin system is ferromagnetic. Indeed, the initial state in Eq.~(\ref{initial_state}) is an eigenstate if the parameters $t_{so}$ and $U_r$ go large. Therefore, the magnetization is about $-1$ for the large values of $t_{so}$ and $U_r$. The small fluctuations around -1 are shown due to the virtual transitions from the perturbed terms.

The odd- and even-number cases are totally different to 
each other. In even-number cases, the effective two-level dynamics
occurs due to the virtual fluctuations of the perturbation terms $\sum_i{S^+_iS^+_{i+1}}+{\rm H.c.}$.
However, these perturbation terms alter the spin state in pair only and therefore they cannot contribute the dynamics
between the two degenerate states $|\uparrow\uparrow\ldots\uparrow\rangle$ and $|\downarrow\downarrow\ldots\downarrow\rangle$ for odd-number cases. Also,
the DM terms cancel the contributions from the spin states and their reflection states. For example,
the states $|001\rangle$ and its reflection states $|100
\rangle$ will be cancelled in the perturbation series.
Therefore, the DM terms cannot lead to the effective
two-level dynamics in this case.

\section{Discussion}
We have investigated the non-equilibrium dynamics of 
a closed quantum system by suddenly applying the SO coupling. It is necessary to produce and tune the required SO couplings to the atoms. There are several
ways to create SO coupling such as two-photon Raman
transition \cite{Liu0,Lin} and shaking lattice \cite{Sengstock,Goldman}. Recently, the techniques
for tuning SO coupling \cite{Garcia} have been demonstrated by using
Raman coupling with laser fields. However, this method
may produce unwanted heating due to spontaneous emissions of atoms. 

Alternatively, the SO coupling can be
exploited by shaking the lattice periodically. This ``shaking'' method has been used to successfully generate
the artificial gauge potential to cold atoms in lattices \cite{Struck}. More
recently, the theoretical proposals for the realization of SO coupling have been put forward \cite{Sengstock,Goldman}. This method is able to
create SO coupling without heating if the appropriate 
driving conditions are met \cite{Sengstock}. The other scheme, which 
overcomes the problems of spontaneous emissions, has also
been proposed \cite{Kennedy}. Apart from that, it is also required to adjust the inter- and intra-component interaction strengths for observing the different kinds of dynamics. This can be made by using Feshbach resonance \cite{Chin}. The scattering length between the different components of atoms can be modified by applying the appropriate magnetic fields to the atoms \cite{Erhard}.

In addition, we make a rough estimation of the relaxation time-scale in realistic experiments. We take the typical value of the tunnel coupling $J$ to be about $100{\sim}200$ Hz in optical-lattice experiments\cite{Fukuhara,Preiss}. Since the ratio $U_{ab}/J$ is tunable \cite{Greiner}, we assume that it ranges between 5 to 10 to enter the Mott insulator regime. This gives the parameter $\lambda=2J^2/U_{ab}$ to be roughly about $40$ to $80$ Hz. In Fig.~\ref{relax_magnet}, we can see that the system takes the period $\lambda$ for relaxation which is about $10\sim{20}$ ms. To observe the intrinsic effect of relaxation in an isolated system, the relaxation time must be much shorter than the damping time from the external noise sources. The typical heating time of atoms in optical lattice is several hundreds of ms \cite{Fukuhara}. This suggests that the relaxation of local spins can be detected in experiments. To take the time-average of magnetization, the required time is about $10\lambda\sim{100}$ ms which is comparable to the heating time. In realistic experiments, the average time of extracting the magnetization can be chosen to be shorter than the heating time to characterize the dynamical properties of the system.

\section{Conclusion}
In summary, we have studied the many-body dynamics 
of two-component bosonic atoms in a 1D optical lattice
by suddenly introducing the SO coupling. In Mott-insulating regime, the system can be described by
a quantum spin system. We study the 
dynamics of magnetization of the system. We find that
the time average of magnetization is useful for characterizing the non-equilibrium dynamics. The system
shows the relaxation and non-stationary dynamics
in the different interaction regimes.
In the relaxation regime, the magnetization becomes nearly stationary in a long time and the local
spins become nearly fully relaxed. When 
the SO coupling is strong and the inter-component interaction strength is sufficiently larger than the intra-component strength, the system becomes non-stationary.
The totally different dynamical behaviours are shown
for the even and odd numbers of sites in the non-stationary regime. 

\appendix
\section{Derivation of effective Rabi frequency}
We study the effective Rabi frequency between the two degenerate states from the perturbation theory. We consider the pair-excitation interaction $V$ to be perturbation. The Hamiltonian $H^s_{\rm eff}\approx{H_0}+V$, where $H_0$ and $V$ are given by
\begin{eqnarray}
H_0&=&\lambda\sum^{N-1}_{i=1}\Big\{2\Big[\Big(\frac{t_{so}}{J}\Big)^2-1\Big]\Big(2\frac{U_{ab}}{U}-1\Big)S^z_iS^z_{i+1},\nonumber\\
V&=&\lambda\Big(\frac{t_{so}}{J}\Big)^2(S^+_iS^+_{i+1}+S^{-}_iS^-_{i+1}),
\end{eqnarray}
where $H_0$ is treated as an unperturbed Hamiltonian
and $V$ as a perturbation. 

To calculate the effective Rabi frequency, we need to evaluate the virtual transition from $|\uparrow\uparrow\ldots\uparrow\rangle$ and $|\downarrow\downarrow\ldots\downarrow\rangle$. The perturbation term $V$ can make the transitions for two neighboring spins in pair. It takes $N/2$ virtual transitions from $|\uparrow\uparrow\ldots\uparrow\rangle$ and $|\downarrow\downarrow\ldots\downarrow\rangle$ only. We can then make the approximation by using the $N/2$-th order perturbation theory. However, the DM terms will take $N$ virtual transitions to connect these two states. The $N$-th order perturbation theory has to be used. The correction from the DM terms is much smaller than that from the term $V$. Therefore, we can safely ignore the DM terms in calculating the perturbation theory if $t_{so}/J$ is comparable with $U_{ab}/U$.

We can obtain the leading terms of the $N/2$-th order eigenenergy:
\begin{eqnarray}
E^{(N/2)}_{n_1,n_2}&=&\sum^{\tilde{N}_c}_{j=1}\frac{V_{n_2k_{N/2-1}}(\prod^{N/2-1}_{i=1}{V^j_{k_{i+1}k_{i}}})V^j_{k_{1}n_1}}{\prod^{N/2-1}_{i=1}E^j_{l_Dk_{i}}}\nonumber\\
&&+{\rm other~terms},
\end{eqnarray}
where $V^j_{lk}={}_{j}\langle{l}|V|{k}\rangle_j$, 
$E^j_{lk}=E^{(0)}_{l_D}-E^{(0)}_{k_j}$ and $D$ denotes the
degenerate subspace for $n_1=|\uparrow\uparrow\ldots\uparrow\rangle$ and $n_2=|\downarrow\downarrow\ldots\downarrow\rangle$, and $\tilde{N}_c$ is the number of possible terms that connect the two degenerate states via virtual fluctuations. The number $\tilde{N}_c$ can be obtained numerically by counting all possibilities to connect the two states. The leading term of $E^{(N/2)}_{n_1,n_2}$ can be written as
\begin{eqnarray}
E^{(N/2)}_{n_1,n_2}&\approx&\lambda\Big(\frac{t_{so}}{J}\Big)2^{1-N/2}(2U_r-1)^{1-N/2}\tilde{N}_c.
\end{eqnarray}

We compare the effective Rabi frequency between
the numerics and the approximation from the perturbation theory in Table \ref{table}, for the different sizes $N$.
The percent error $\eta$ shows the error between 
the exact numerical value and the approximation. It
is defined as
\begin{eqnarray}
\eta&=&\frac{|\Omega_R-\tilde{\Omega}_R|}{|\Omega_R|}\times{100}\%.
\end{eqnarray}
Here we denote $\tilde{\Omega}_R=E^{(N/2)}_{n_1,n_2}$ as the approximation from the perturbation theory. 
This approximation is fairly good when $N$ is small. 
However, as $N$ increases, the error grows. In fact, we
have taken account of the leading term from the $N/2$-th
order perturbation only. When $N$ increases, the
calculation should include the higher order perturbation terms to improve the accuracy.

\begin{table}
\begin{center}
\caption{This table shows the effective Rabi frequencies from the numerical results and the perturbation theory for the different system's sizes, and the percent error $\eta$.}
\label{table}
\begin{tabular}{|c|c|c|c|}
\hline 
$N$ & $\Omega_R/\lambda$ & $\tilde{\Omega}_R/\lambda$ & $\eta$\\ 
\hline 
6 & 8.9760 & 8.1633 & 9.1\% \\ 
\hline 
8 & 2.7013 & 2.3324 & 13.7\%\\ 
\hline 
10 & 0.83642 & 0.66639 & 20.3\%\\ 
\hline 
12 & 0.25964  & 0.19040 & 26.7\%\\ 
\hline 
\end{tabular} 
\end{center}
\end{table}

\begin{acknowledgments}
This work was supported in part by the 
National Basic Research Program of 
China Grant No. 2011CBA00300 and No. 2011CBA00301, 
and the National Natural Science Foundation of 
China Grant No.11304178, No. 61061130540, and No.
61361136003.
\end{acknowledgments}

\appendix

\end{document}